\documentstyle[12pt]{article}
\input epsf
\catcode`\@=11
\def\marginnote#1{}

\newcount\hour
\newcount\minute
\newtoks\amorpm
\hour=\time\divide\hour by60
\minute=\time{\multiply\hour by60 \global\advance\minute by-
\hour}
\edef\standardtime{{\ifnum\hour<12 \global\amorpm={am}%
    \else\global\amorpm={pm}\advance\hour by-12 \fi
    \ifnum\hour=0 \hour=12 \fi
    \number\hour:\ifnum\minute<100\fi\number\minute\the\amorpm}}
\edef\militarytime{\number\hour:\ifnum\minute<100\fi\number\minute}
\def\draftlabel#1{{\@bsphack\if@filesw {\let\thepage\relax
  \xdef\@gtempa{\write\@auxout{\string
    \newlabel{#1}{{\@currentlabel}{\thepage}}}}}\@gtempa
    \if@nobreak \ifvmode\nobreak\fi\fi\fi\@esphack}
     \gdef\@eqnlabel{#1}}
\def\@eqnlabel{}
\def\@vacuum{}
\def\draftmarginnote#1{\marginpar{\raggedright\scriptsize\tt#1}}
\def\draft{\oddsidemargin -.5truein
        \def\@oddfoot{\sl preliminary draft \hfil
        \rm\thepage\hfil\sl\today\quad\militarytime}
        \let\@evenfoot\@oddfoot \overfullrule 3pt
        \let\label=\draftlabel
        \let\marginnote=\draftmarginnote

\def\@eqnnum{(\theequation)\rlap{\kern\marginparsep\tt\@eqnlabel}%
\global\let\@eqnlabel\@vacuum}  }
\def\preprint{\twocolumn\sloppy\flushbottom\parindent 1em
        \leftmargini 2em\leftmarginv .5em\leftmarginvi .5em
        \oddsidemargin -.5in    \evensidemargin -.5in
        \columnsep 15mm \footheight 0pt
        \textwidth 250mmin      \topmargin  -.4in
        \headheight 12pt \topskip .4in
        \textheight 175mm
        \footskip 0pt

\def\@oddhead{\thepage\hfil\addtocounter{page}{1}\thepage}
        \let\@evenhead\@oddhead \def\@oddfoot{} \def\@evenfoot{}
}
\def\titlepage{\@restonecolfalse\if@twocolumn\@restonecoltrue\o
necolumn
     \else \newpage \fi \thispagestyle{empty}\c@page\z@
        \def\thefootnote{\fnsymbol{footnote}} }
\def\endtitlepage{\if@restonecol\twocolumn \else  \fi
        \def\thefootnote{\arabic{footnote}}
        \setcounter{footnote}{0}}  
\catcode`@=12
\relax
\def\singlespaced{\baselineskip=\normalbaselineskip}

\def\tt{\tilde t}

\newcommand{\newc}{\newcommand}
\newcommand\eg{{\it {e.g.}}}
\newcommand\etal{{\it {et al.}}}
\newcommand\lsim{\mathrel{\rlap{\lower4pt\hbox{\hskip1pt$\sim$}}
    \raise1pt\hbox{$<$}}}
\newcommand\gsim{\mathrel{\rlap{\lower4pt\hbox{\hskip1pt$\sim$}}
    \raise1pt\hbox{$>$}}}

\newcommand\mchi{m_{\chi}}              

\newc{\sthw}{\sin\theta_W}              \newc{\cthw}{\cos\theta_W}
\newc{\bino}{\widetilde B}              \newc{\wino}{\widetilde W_3}
\newc{\higgsinob}{{\widetilde H}^0_b}   \newc{\higgsinot}{{\widetilde H}^0_t}

\newc{\abund}{\Omega h^2}
\newc{\abundchi}{\Omega_\chi h^2}
\newc{\rhocrit}{\rho_{crit}}
\newc{\rhochi}{\rho_{\chi}}

\newc{\mwimp}{m_{\chi}}     \newc{\rhowimp}{\rho_{\chi}}

\newc{\mplanck}{M_{\rm P}}              \newc{\mgut}{M_{\rm GUT}}
\newc{\mz}{m_{Z}}                       \newc{\mw}{m_{W}}

\newcommand\gev{\,\mbox{GeV}}

\newcommand\kev{\,\mbox{keV}}

\newcommand\kmpersec{\,\mbox{km/s}}
\newcommand\yr{\,\mbox{yr}}
\newcommand\kg{\,\mbox{kg}}
\newcommand\kgday{\,\mbox{kg$\times$day}}
\newcommand\pb{\,\mbox{pb}}
\newcommand\gevcmcube{\,\mbox{GeV/cm$^3$}}

\newc{\ra}{\rightarrow}
\newc{\beq}{\begin{equation}}
\newc{\eeq}{\end{equation}}
\newc{\bea}{\begin{eqnarray}}
\newc{\eea}{\end{eqnarray}}

\newc{\vearth}{v_{\otimes}}
\newc{\vsun}{v_{\odot}}
\newc{\vsube}{v_e}	\newc{\vzero}{v_0}	\newc{\vrms}{v_{\rm rms}}

\newc{\sigmap}{\sigma_p}
\newc{\rhozerothree}{\rho_{0.3}}

\newc{\sigchin}{\sigma(\chi N)}
\newc{\etazero}{\eta_0}		\newc{\deltaeta}{\Delta\eta}
\newc{\erecoil}{E_R}
\include{psfig}
\begin{document}
\topmargin-1.cm
%
\begin{titlepage}
\vspace*{-64pt}
\begin{flushright}
{UM-TH-99-02\\
hep-ph/9903468\\
Spring Equinox, 1999}\\
\end{flushright}

\vspace{1.8cm}

\begin{center}
\large {\bf WIMP Velocity Impact on Direct Dark Matter Searches} \\
\vspace*{1.3cm}
\large{Michal Brhlik$^a$ and Leszek Roszkowski$^b$} \\
\vspace*{0.4cm}
{\it $^a$Randall Physics Lab }\\
{\it University of Michigan}\\
{\it Ann Arbor, MI 48109-1120, USA}\\
\vspace*{0.4cm}
{\it $^b$Department of Physics}\\
{\it Lancaster University}\\
{\it Lancaster LA1 4YB, UK} 

\end{center}

\bigskip
\begin{abstract}

We examine the effect of some uncertainties in the input astrophysical
parameters on direct detection searches for WIMPs in the Galactic
halo. We concentrate on the possible WIMP annual modulation signal
recently reported by the DAMA Collaboration. 
We find that allowing for a
reasonable uncertainty in a WIMP Maxwellian velocity distribution
leads to significantly relaxed constraints on the WIMP mass as
compared to the original DAMA analysis.

\end{abstract}

\end{titlepage}

\setcounter{footnote}{0}
\setcounter{page}{0}
\newpage


\section{Introduction}

It is generally assumed that spiral galaxies are surrounded by
extended halos of dark matter (DM) for which a likely candidate is a
hypothetical stable and neutral particle generically called a weakly
interacting massive particle (WIMP). Experimental sensitivity has now
improved to the level at which a signal from well-motivated WIMP
candidates, like the neutralino of supersymmetry (SUSY), could be
detected~\cite{jkg}.  However, inherent astrophysical and other
uncertainties are likely to considerably weaken experimental limits and
regions of a possible WIMP signal. 

One of the most promising ways to look for WIMPs
in the halo of the Milky Way is the one based on annual
modulation~\cite{annualmodulation86,annualmodulation88}.  A WIMP
signal rate depends on the effective velocity of an incident WIMP
on a detector target which in turn depends on the Earth's velocity
$\vsube$ in the Galactic frame and on the WIMP velocity distribution
in the halo.  The Earth's motion in the Galactic frame is a
superposition of its rotation around the Sun with that of the Sun
around the Galactic center,
\beq \vsube(t) =
\vsun + \vearth \cos\gamma \cos\left( {{2\pi (t-t_p)}\over{T}} \right)
\label{vsube:eq}
\eeq
where $\vsun\approx232\kmpersec$ and $\vearth\approx30\kmpersec$ are
the respective rotational velocities of the Sun and the Earth,
$\gamma=60^\circ$ is the angle between the two planes of rotation,
$T=1\yr$ and the peak occurs on $t_p=153\,{\rm rd}$ day of the year
(around the 2nd of June).  Although the effect is expected to be
small, around a few per cent, with enough sensitivity it is
measurable.

The annual modulation method has been adopted by the DAMA
Collaboration in a Gran Sasso Laboratory-based experiment involving a
detector consisting of nine $9.70\kg$ NaI(Tl) crystals.  Based on the
statistics of $14,962\kgday$ of data collected over a period from
November '96 to July '97 (part of run~II), the Collaboration has
recently reported~\cite{dama98two} a statistically significant effect
which could be caused by an annual modulation signal due to a WIMP
with mass $\mwimp$ and WIMP-proton cross section $\sigmap$ given as
\beq
\mwimp=59\gev^{+17}_{-14}\gev \ \ \ \ \ \ \ \ \ 
\ \  \xi\sigmap= 7.0^{+0.4}_{-
1.2}\times10^{-6}\pb\ \ \ \ ({\rm at\ 99.6\%\ CL}),
\label{damarange:eq}
\eeq
where $\xi=\rhowimp/\rhozerothree$ stands for the local WIMP mass
density $\rhowimp$ normalized to $\rhozerothree=0.3\gevcmcube$. (See
also Figure~6 in Ref.~\cite{dama98two} for a $2\sigma$ signal region in the
($\mwimp,\xi\sigmap$) plane.)
According to DAMA, the new analysis is consistent with and confirms
the Collaboration's earlier hint~\cite{dama98one} for the presence of the
signal based on $4,549\kgday$ of data.

In this Letter, we point out that the region selected by DAMA as
corresponding to a possible WIMP signal strongly depends on the
assumed values of astrophysical parameters which are poorly known.
Given the fact that the Galactic halo has not been directly observed,
nor any of its parameters have been actually measured, we advocate a
more conservative approach. We demonstrate the effect by varying the
position of the peak of the WIMP velocity distribution which,
following DAMA and a standard lore, we assume here to be Maxwellian.
We find that significantly larger ranges of especially WIMP mass and,
to a lesser extent, $\xi\sigmap$ should be considered as consistent
with the possible annual modulation effect reported by DAMA. This will
have important implications for the allowed configurations of SUSY
masses and couplings.

In Section~2 we summarize the relevant elements of the DAMA
analysis. In Section~3 we present the method which we use for
calculating the expected event rate from Majorana WIMPs in annual
modulation. Results and conclusions are presented in Section~4.

\section{DAMA Analysis}

The expected WIMP signal event rate (see Section~3 for details) 
is proportional to the elastic
WIMP-nucleus cross section $\sigchin$, the local WIMP number density
$\rho_\chi/\mchi$ and the density of the target nuclei with mass
$m_A$. The local WIMP density enters multiplicatively and it is
possible to account for its variation by normalizing it to some
nominal value, \eg, $\rhozerothree$ in Ref.~\cite{dama98two}, and
multiplying the scattering cross section by $\xi$.
The event rate also depends on the  velocity of a halo WIMP
relative to the detector but this dependence is highly non-trivial and
cannot be factorized out.

Assuming the WIMP velocity distribution to be characterized by a parameter
$\vzero$, one can re-write Eq.~(\ref{vsube:eq}) as
\beq
\eta(t) = \etazero + \deltaeta \cos\left( {{2\pi (t-t_p)}\over{T}}
\right)
\label{eta:eq}
\eeq
where $\eta=\vsube/\vzero$, $\eta_0=\vsun/\vzero$, and $\deltaeta=
\vearth \cos\gamma /\vzero$. Since $\vsun\gg\vearth$ and it is expected that
$\vzero\sim\vsun$, it is reasonable to assume that
$\etazero\gg\deltaeta$. This was used by DAMA to approximate the
differential 
signal event
rate in the k-th energy bin by the first two terms
in Taylor's expansion~\cite{dama98one}
\bea
S_k\left[\eta(t)\right] &=& S_k\left[\etazero\right] 
+ \left[\frac{\partial S_k}{\partial\eta} \right]_{\etazero} 
\deltaeta \cos\left( {{2\pi (t-t_p)}\over{T}}
\right)\cr
&=& S_{0,k} + S_{m,k}\cos\left( {{2\pi (t-t_p)}\over{T}}
\right).
\label{taylor:eq}
\eea

Using a maximum likelihood method~\cite{dama98two,dama98one}, the
(time-independent) background was separated from the constant component
$S_{0,k}$ and the time-dependent component $S_{m,k}$.  Because of the
difference in the profiles of the expected energy distributions, the
largest contribution to the signal was expected from the lowest energy
bins above the threshold~\cite{dama98one}.

Based on the statistics of $14,962\kgday$ of data,
values of $S_{0,k}$ and $S_{m,k}$ were obtained between $2\kev$ 
(the software
threshold) and $20\kev$ as shown in
the Table~\cite{dama98two}.
\vspace{0.3cm}
\begin{center}
\begin{tabular}{|c|c|c|} \hline
Energy & $S_{0,k} \pm \sigma_{0,k}$ & $S_{m,k} \pm \sigma_{m,k}$ \\
(keV)  & (\rm{cpd/kg/keV}) & (\rm{cpd/kg/keV}) \\ 
\hline \hline
2 - 3 & $0.54\pm 0.15$ & $0.018\pm 0.009$ \\ \hline
3 - 4 & $0.23\pm 0.08$ & $0.012\pm 0.004$ \\ \hline
4 - 5 & $0.09\pm 0.04$ & $0.006\pm 0.002$ \\ \hline
5 - 6 & $0.04\pm 0.02$ & $0.003\pm 0.001$ \\ \hline
\end{tabular}
\end{center}   
\vspace{0.3cm}
Above 6 keV the detected time-dependent component is statistically 
insignificant. 

As mentioned above, it is possible to write $S_{0,k}= \xi\sigmap
S^{'}_{0,k}$ and $S_{m,k}= \xi\sigmap
S^{'}_{m,k}$~\cite{dama98one}. Both $S^{'}_{0,k}$ and $S^{'}_{m,k}$
depend on WIMP mass but also on its velocity distribution, in addition
to several other detector and nuclear factors. Expressions for
$S^{'}_{0,k}$ and $S^{'}_{m,k}$ used by DAMA can be found in
Ref.~\cite{dama96}. The ranges~(\ref{damarange:eq}) of $\mwimp$ and
$\xi\sigmap$ in agreement with the hypothesis of an annual modulation
signal due to a Majorana WIMP with dominant spin independent interaction
with NaI target were obtained by a maximum likelihood
method~\cite{dama98two}. The period was assumed to be one year and the
position of the peak was fixed at the expected value.

We note that in DAMA's analysis a spherical halo model and a
Maxwellian velocity distribution with a {\em fixed} value of $\vzero=
\sqrt{2/3}\vrms=220\kmpersec$ was assumed~\cite{dama98two}. We will
argue that the ranges of $\mwimp$ and $\xi\sigmap$ depend sensitively
on the assumed value of $\vzero$. First we will present the procedure
which we use.

\section{Majorana WIMP Detection Rate in NaI Detectors}

Elastic interactions of relic WIMPs with nuclei in the detector result
in a nuclear recoil energy deposit in the detector volume. Their size
depends on the cross section of the WIMP scattering off constituent
quarks and gluons.  For non-relativistic Majorana particles, these can
be divided into two separate types~\cite{gw}. 
The coherent part described by an effective scalar
coupling between the WIMP and the nucleus is proportional to the
number of nucleons in the nucleus. It receives a tree-level contribution from
scattering off quarks, $\chi q\ra \chi q$, as described by a Lagrangian ${\cal
L}\sim \left(\chi\chi\right)\left(\bar q q\right)$.
The incoherent component of the WIMP-nucleus cross section results
from an axial current interaction of a WIMP with constituent quarks,
given by 
${\cal L}\sim \left(\chi\gamma^\mu\gamma_5\chi\right)\left(\bar
q\gamma_\mu\gamma_5 q\right)$, and couples the spin of the WIMP to the
total spin of the nucleus.

In the case of a supersymmetric neutralino, there are several diagrams 
contributing to the scalar part, mainly from Higgs exchange and squark
exchange  with, in general, non-degenerate left and right squark soft masses.
Another contribution to coherent interactions comes from
one-loop neutralino-gluon scattering where the exchanged Higgs couples
to a heavy quark loop on a gluon line~\cite{drees}. The axial interaction is 
due to the $Z$ and squark exchange.

The differential cross section for a WIMP scattering
off a nucleus $X_{Z}^{A}$ with mass $m_A$ is therefore expressed as 
\begin{eqnarray}
\frac{d\sigma}{d|\vec{q}|^2}=\frac{d\sigma^{scalar}}{d|\vec{q}|^2}+
\frac{\ d\sigma^{axial\ }}{d|\vec{q}|^2},
\label{signucleus:eq}
\end{eqnarray}
where the transferred momentum $\vec{q}=\frac{m_A
\mwimp}{m_A+\mwimp}\vec{v}$ depends on the velocity $\vec{v}$ of the
incident WIMP.  The effective WIMP-nucleon cross sections
$\sigma^{scalar}$ and $\sigma^{axial}$ are computed by evaluating
nucleonic matrix elements of corresponding WIMP-quark and WIMP-gluon
interaction operators.  

In the scalar part contributions from individual
nucleons in the nucleus add coherently and the finite size effects are
accounted for by including the scalar nuclear form factor $F(q)$.
(The effective interaction in general also
includes tensor components but the relevant nucleonic matrix elements
can be expanded in the low momentum-transfer limit in terms of the
nucleon four-momentum and the quark (gluon) parton distribution
function.  As a result, the non-relativistic WIMP-nucleon Lagrangian
contains only scalar interaction terms.) 
The differential cross section for the scalar part then takes the form 
\cite{jkg}     
\begin{eqnarray}
\frac{d\sigma^{scalar}}{d|\vec{q}|^2}=\frac{1}{\pi v^2}[Z f_p +(A-Z) f_n]^2
F^2 (q),
\end{eqnarray}
where $f_{p}$ and $f_{n}$ are the effective neutralino couplings to
protons and neutrons, respectively. Explicit expressions for the case
of the supersymmetric neutralino can be found in the Appendix of 
Ref.~\cite{bb}.

The effective axial
WIMP  coupling to the nucleus  depends on the spin content
of the nucleon $\Delta q_{p,n}$ and the overall expectation value of the 
nucleon group spin in the nucleus $<S_{p,n}>$. For a nucleus with a total
angular momentum $J$ we have
\begin{eqnarray}
\frac{d\sigma^{axial}}{d|\vec{q}|^2}=\frac{8}{\pi v^2}\Lambda^2 J (J+1)
S(q),
\end{eqnarray}
with $\Lambda=\frac{1}{J} [a_p \langle S_p\rangle+a_n \langle S_n\rangle]$.
The axial couplings 
\begin{eqnarray}
a_p=\frac{1}{\sqrt{2}} \sum_{\scriptstyle u,d,s} d_q \Delta q^{(p)},
&\ \ \  &
a_n=\frac{1}{\sqrt{2}} \sum_{\scriptstyle u,d,s} d_q \Delta q^{(n)}
\end{eqnarray}
are determined by the experimental values of the spin constants $\Delta 
u^{(p)}=\Delta d^{(n)}=0.78$, $\Delta d^{(p)}=\Delta u^{(n)}=-0.5$ and 
$\Delta s^{(p)}=
\Delta s^{(n)}=-0.16$. 
The effective couplings $d_q$ depend on the WIMP properties and for
the neutralino they can be found in the Appendix of Ref.~\cite{bb}.

In translating $\sigma(\chi q)$ and $\sigma(\chi g)$ into the
WIMP-nucleon cross section in~Eq.~(\ref{signucleus:eq}) several
uncertainties arise. The nucleonic matrix element coefficients for the scalar
interaction are not precisely known. Also, the spin content of the nucleon
and the expectation values of the proton (neutron) group spin in a particular 
nucleus are fraught with significant uncertainty and nuclear model 
dependence. These ambiguities have to be considered in numerical calculations.
Finally, in order to obtain $\sigma(\chi N)$, models of nuclear
wave functions must be used. The scalar nuclear
form factor reflects  the mass density density distribution in the nucleus.
Following DAMA analysis we take the form factor for Na to be 
equal to one while for I we use the Saxon-Woods form factor \cite{engel}
\begin{eqnarray}
F(q)=\frac{3j_1 (qR_0)}{qR_1}
e^{-\frac{1}{2}(qs)^2},
\end{eqnarray}
where $R_1=\sqrt{R^2-5s^2}$, $R=A^{\frac{1}{3}} \times 1.2\,\rm fm$, $j_1$ is a
spherical Bessel function and $s=1\,\rm fm$. 

The spin form factor for I is assumed to be 
\begin{eqnarray}  
S(q)=[0.6845 e^{-359.1 q^2}-3.427 q^2+0.3042] e^{-0.2 q^2 (r_I^2-r_{Xe}^2)}
\end{eqnarray}
where $q$ is in GeV and the nuclear radius is $r=1.2 A^{1/3}\, {\rm fm}$.
The finite size effects for Na can be neglected. The nucleon group spin 
expectation values for I have been estimated within the quenched interacting
boson model giving $<S_p>=0.128$ and $<S_n>=0$ \cite{io}, and for Na we use
the odd group model result $<S_p>=0.136$ and $<S_n>=0$ \cite{na}. 

In order to calculate the WIMP detection rate for a given  material,
it is necessary to convolute the  cross
section~Eq.~(\ref{signucleus:eq}) with the local WIMP flux which, in
the rest frame of the detector, will be time-dependent.  The
differential event rate can be expressed as~\cite{jkg}
\begin{eqnarray}
\frac{dR}{d\erecoil}&=&\frac{4}{\sqrt{\pi^3}}\frac{\rhowimp}{\mwimp} 
\widetilde T(\erecoil)
\biggl \{ [Z f_p +(A-Z) f_n]^2 F^2 (q)\cr
& &\ \ \ \ \ \ + 8\Lambda^2 J (J+1) S(q) \biggr\},
\label{eventrate:eq}
\end{eqnarray}  
where $\erecoil=\frac{|\vec{q}|^2}{2m_A}$ is the recoil energy of the 
nucleus. The time-dependent function $\widetilde T(\erecoil)$
integrates over all possible kinematic configurations 
in the scattering process 
\begin{eqnarray}
\widetilde T(\erecoil)=\frac{\sqrt{\pi}}{2}\int_{v_{min}}^{\infty}
\frac{f_{\chi}}{v}\, dv.
\label{tdef:eq}
\end{eqnarray}

For a Maxwellian velocity distribution taking into account the motion of the 
Sun and the Earth, Eq.~(\ref{vsube:eq}), one obtains
\begin{eqnarray}
\widetilde T(\erecoil)=\frac{\sqrt{\pi}}{4v_e}
\left[
Erf\left(\frac{v_{min}+v_e(t)}{v_0}\right)-
Erf\left(\frac{v_{min}-v_e(t)}{v_0}\right)\right],
\label{tmaxwell:eq}
\end{eqnarray}
with $v_{min}=\sqrt{\frac{\erecoil (\mwimp+m_A)^2}{2 m^2_{\chi} m_A}}$.

The actually measured response from a nucleus $X$ hit by a WIMP
is only a fraction of the recoil energy, $E=q_{X}\erecoil$. It is also
called the electron equivalent energy and is determined by the
quenching factor $q_{X}$ which is different for each detector
material. Here we assume numerical values of the quenching factors
$q_{Na}=0.30$ and $q_{I}=0.09$ as they were reported by the DAMA
Collaboration~\cite{dama96} in agreement with previously published
results~\cite{quench}.

The expected experimental spectrum per energy bin can then be
expressed as
\begin{eqnarray}
\frac{\Delta R}{\Delta E}(E)=r_{Na}\int_{E/q_{Na}}^{(E+\Delta E)/q_{Na}}
\frac{dR_{Na}}{d\erecoil}(\erecoil)\, \frac{d\erecoil}{\Delta E}+
r_{I}\int_{E/q_{I}}^{(E+\Delta E)/q_{I}}
\frac{dR_{I}}{d\erecoil}(\erecoil)\, \frac{d\erecoil}{\Delta E},
\label{spectrum:eq}
\end{eqnarray}
where $r_{Na}=\frac{M_{Na}}{M_{Na}+M_{I}}=0.153$ and
$r_{I}=\frac{M_{I}}{M_{Na}+M_{I}}=0.847$ are the respective mass fractions
of Na and I in the detector material. 

The formulae presented here are in full agreement with those used by
DAMA~\cite{dama96}. The fact that we neglect the WIMP escape velocity
from the halo (estimated in Ref.~\cite{annualmodulation86} between 
$580\kmpersec$ and $625\kmpersec$)
leads to a negligible numerical difference. The differential event
rate in Eq.~(\ref{eventrate:eq}) is dominated by the scalar
contribution due to WIMP scattering off iodine.  In our
calculation we assume that $f_p\approx f_n$ \cite{jkg} and use $f_p$
as a free parameter which can be directly translated into the WIMP
cross section on proton $\sigma_p=\frac{4 m_p^2 m_{\chi}^2}{\pi
(m_p+m_{\chi})^2} f_p^2$.  In this approximation one can re-write
Eq.~(\ref{eventrate:eq}) as 
\begin{eqnarray}
\frac{dR}{d\erecoil}=
\left(\xi\sigmap\right)\left[\frac{\rhozerothree}{\sqrt{\pi}}
\frac{\left(m_p+\mwimp\right)^2}{m_p^2\mwimp^3} A^2 {\widetilde
T}(\erecoil) F^2(q).
\right]
\label{rateapprox:eq}
\end{eqnarray}
Theoretical prediction for the signal is then calculated using
Eqs.~(\ref{eventrate:eq}) and~(\ref{spectrum:eq}) with $\Delta
E=1\,{\rm keV}$.  The predicted rates in each channel are evaluated for
$t=t_p$ and $t=t_p+T/2$, and Eq.~(\ref{taylor:eq})
is used to extract the predicted values 
$S_{0,k}^{th}$ and
$S_{m,k}^{th}$. These are compared with the experimental values given
in the Table, as described below.

\section{Results and Conclusions}

Our goal is to examine how the sensitivity of the expected event
spectrum will depend on the velocities of incident WIMPs. It is clear
that these enter in a rather complicated way, through
Eq.~(\ref{tmaxwell:eq}), but also through the transferred momentum $q$
on which the form factors depend.

Experimental determinations of the Galactic halo velocity
distributions are rather uncertain. Assuming a spherically symmetric
and isotropic halo with a Maxwellian dark matter velocity
distribution, it was argued in Ref.~\cite{annualmodulation86} that
$\vzero$ should be of order of the local Galactic rotation
velocity. Several values have been reported for the latter:
$243\pm20\kmpersec$~\cite{rotvel1},
$222\pm20\kmpersec$~\cite{rotvel2},
$228\pm19\kmpersec$~\cite{rotvel3}, although a much larger spread
$220\pm57\kmpersec$ was also recently quoted~\cite{kk98}.  To be
consistent with the DAMA analysis where a fixed value of
$\vzero=220\kmpersec$ was assumed, we take $\vzero=220\pm20\kmpersec$.

\begin{figure}[h!]      
\centering
\epsfxsize=3.75in                
\hspace*{0in}
\epsffile{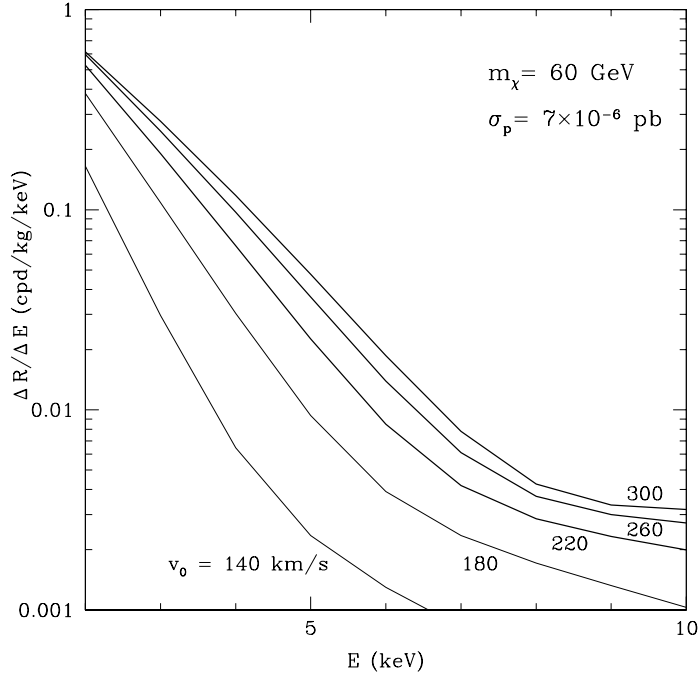}
\bigskip
\caption{The expected differential event rate spectrum versus the detected
energy $E$ for $\mwimp=60\gev$, $\sigmap=7\times10^{-6}\pb$ and several 
choices of $\vzero$.
}
\label{figone}
\end{figure}

We first illustrate in Figure~\ref{figone} the effect of the
dependence of the expected event spectrum on the velocities of
incident WIMP. We plot the expected differential event rate versus the
detected energy $E$ for several choices of $\vzero$. We find a
significant dependence on $\vzero$ at both smaller and larger values
of the detected energy. The effect is caused by the dependence of
$\widetilde T$ on the position of the peak of the WIMP velocity distribution 
$\vzero$. At fixed $\mwimp$ and $\sigma_p$, $\widetilde T(E)$
increases with  $\vzero$ for both sodium and iodine.

The bending appearing in the spectra at lower values of $\frac{\Delta
R}{\Delta E}$ is caused by the contribution from WIMPs scattering off
Na nuclei becoming larger than that off iodine. The latter decreases
more rapidly with energy due to a smaller quenching factor. It is
worth noting that in the future, with large enough statistics
available, the position of the bending, which depends on the WIMP's
mass and velocity, could provide additional information about the
signal.

In order to compare the experimental spectrum measured by DAMA with
theoretical WIMP predictions, we introduce a function
$\kappa$ defined as
\begin{eqnarray}
\kappa=\sum_{\scriptstyle k} \frac{\left(S_{0,k}^{th}-S_{0,k}^{exp}\right)^2}
{\sigma_{0,k}^2}+
\sum_{\scriptstyle k} \frac{\left(S_{m,k}^{th}-S_{m,k}^{exp}\right)^2}
{\sigma_{m,k}^2}
\label{kappa:eq}
\end{eqnarray} 
with the experimental errors of both the time-independent and
time-dependent signal components $\sigma_{0,k}$ and $\sigma_{m,k}$,
respectively, serving as weights. (They are defined in the Table.)
Minimization of this function is then used to look for the ranges of
WIMP masses and cross sections which would best fit the ranges of
$S_{0,k}$ and $S_{m,k}$ responsible for DAMA's possible annual
modulation signal.

We stress that the function $\kappa$ cannot serve as a substitute
for a dedicated maximum likelihood analysis employed by DAMA where
detector efficiencies for each crystal and other factors were included
to evaluate the background and both the signal components $S_{0,k}$
and $S_{m,k}$. Only a full experimental analysis can lead to selecting
a region in the plane ($\mwimp,\xi\sigmap$) consistent with the
signal. Nevertheless, we believe that $\kappa$ is an adequate tool to
demonstrate the dependence of a signal region on WIMP velocities. This
is presented in Figure~2 where we plot contours of $\kappa=10$ for
$\vzero=220\kmpersec$ (solid), which is the value used by DAMA, and
for $\vzero=180\kmpersec$ (long dash) and $\vzero=260\kmpersec$
(short dash) corresponding to a $2\sigma$ error in $\vzero$. We note
that the function $\kappa$ is well-focused and that the contour
$\kappa=10$ reproduces the $2\sigma$ region of DAMA (the dotted curve
in Figure~2 adopted from Figure~6 in Ref.~\cite{dama98two}) remarkably
well.  Any other choice of $\kappa$ below 20 or so
would give only somewhat more relaxed contours.

\begin{figure}[t!]      
\centering
\epsfxsize=3.75in                
\hspace*{0in}
\epsffile{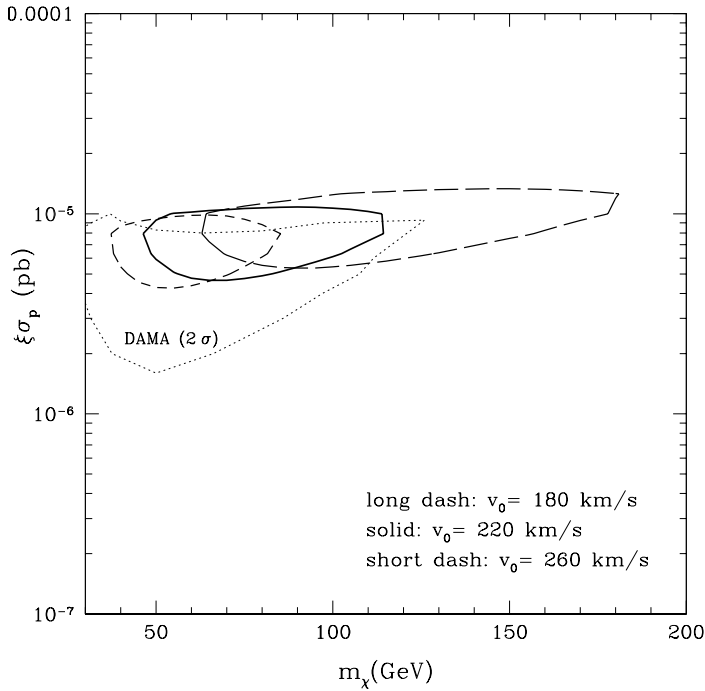}
\bigskip
\caption{Contours of the function
$\kappa=10$~(Eq.~\protect{\ref{kappa:eq}}) for 
different values of the peak of the halo WIMP Maxwellian velocity
distribution. Denoted by dots is the $2\sigma$ region selected by DAMA
assuming $v_0=220\kmpersec$. 
}
\label{figtwo}
\end{figure}

The region of $\kappa=10$ slides quickly to the right and becomes much
more elongated for $\vzero$ decreasing from the value
$\vzero=220\kmpersec$ used in the DAMA analysis~\cite{dama98two}.
We find a significant relaxation of the
upper limit on $\mwimp$ even for values of $\vzero$ much closer to 
$\vzero=220\kmpersec$. For example, for $\vzero=200\kmpersec$,
the region reaches $\mwimp\approx140\gev$.   For $\vzero$ larger than
$220\kmpersec$ the region moves to the left and becomes much more
confined. Its vertical position decreases only very slowly with
increasing velocity.

It is clear from Figure~2 that the dependence of $\kappa$ on $\vzero$
is very strong and cannot be neglected. This effect is particularly
important for smaller WIMP velocities as it allows for substantially
larger values of the WIMP mass to be considered compatible with the
possible annual modulation signal from DAMA, as already emphasized in
Ref.~\cite{cosmo98-lr}. (A crude estimate of the effect was also given
in Ref.~\cite{fornengo-idm98}.)  We find that varying $\vzero$ 
within the $2\sigma$ range around $220\kmpersec$
($180\lsim\vzero\lsim260\kmpersec$) leads to increasing the upper
limit for $\mwimp$ by nearly $60\%$ while on the lower
side of $\mwimp$ the relaxation is only about $25\%$. The range of the
$\xi\sigmap$ values increases by about 10\%.

\begin{figure}[t!]
\centerline{ \epsfxsize 3.15 truein \epsfbox {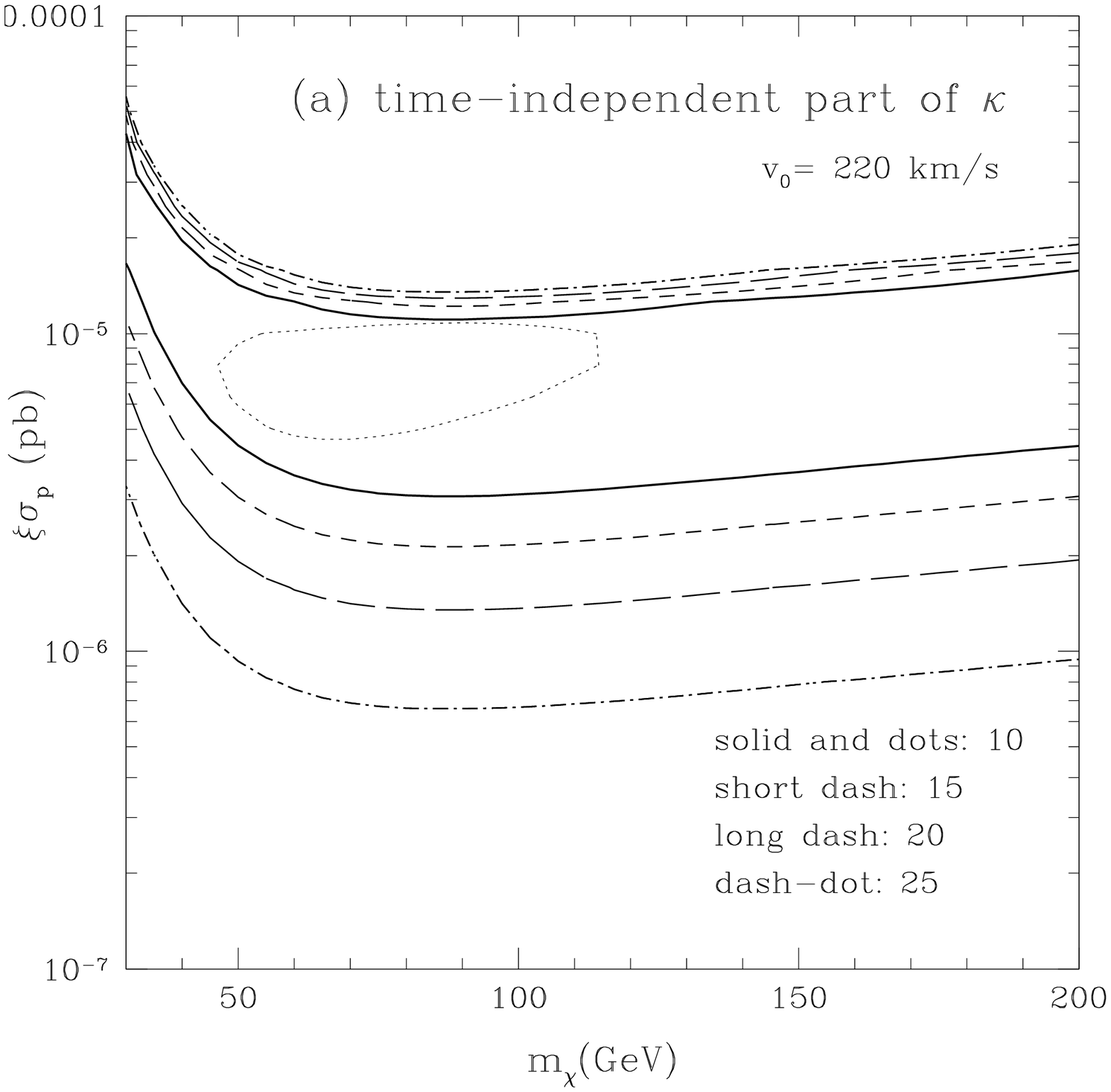}
             \epsfxsize 3.15 truein \epsfbox {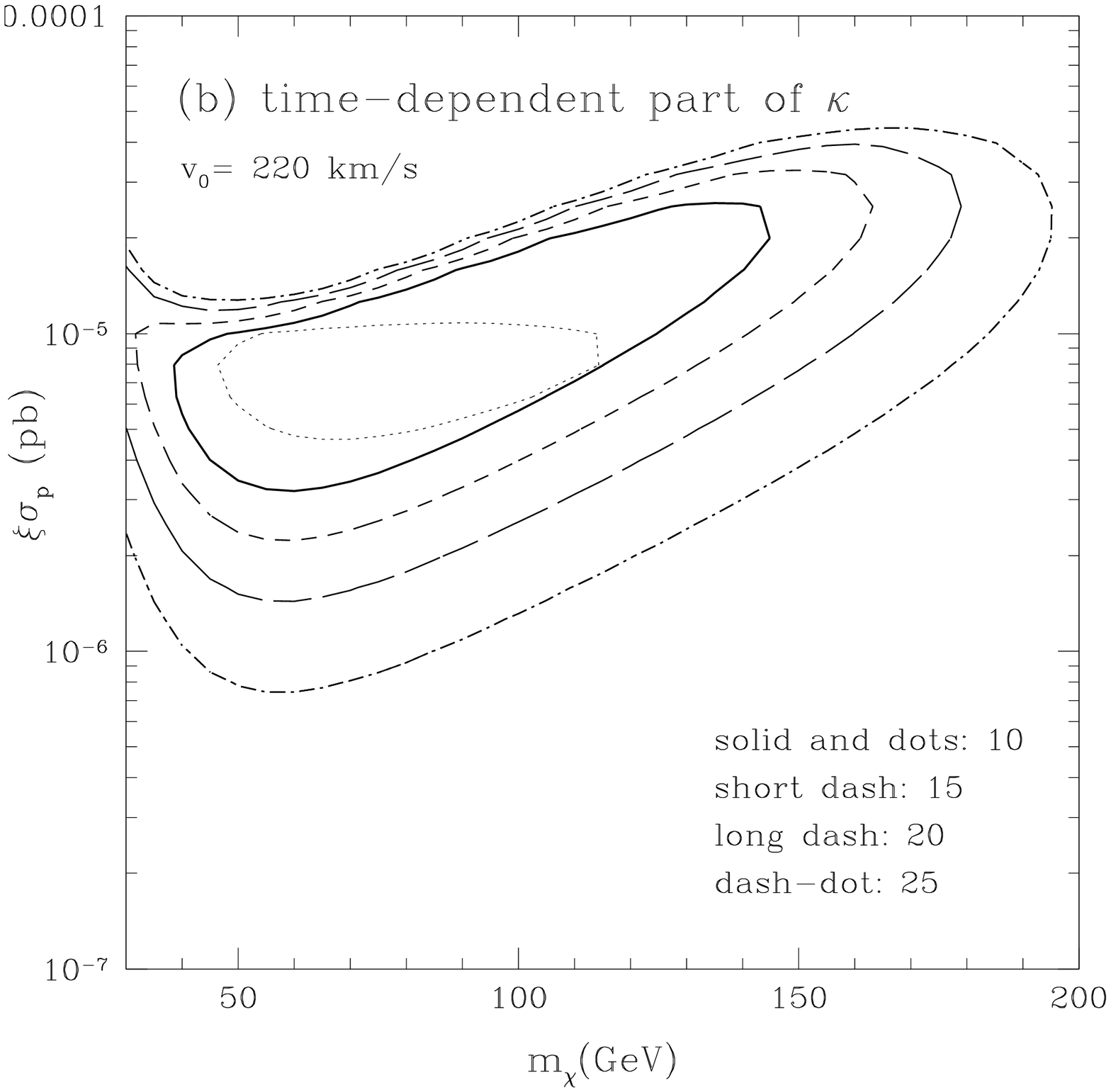}}
\caption{Several contours of (a) the time-independent and
(b) time-dependent 
parts of $\kappa$ for $v_0=220\kmpersec$.
Denoted by dots is the contour of 10 of total $\kappa$.
}
\label{figthree}
\end{figure}
Next we examine the relative contributions from  the  time-independent and
time-dependent parts to the shape of $\kappa$ contours.  This is
shown in Figure~3 for the case $\vzero=220\kmpersec$. We find an
interesting interplay. The time-independent part of $\kappa$ quickly
increases around $1.1\times 10^{-5}\pb$ (except for smaller
$\mwimp\lsim50\gev$) but does not put any restriction on large
$\mwimp$. It is the time-dependent part which cuts off the large WIMP
mass region allowed by the constant component while at the same time
favoring larger values of $\xi\sigmap$ at larger $\mwimp$. With more
data taken over a full  period or more available, this property may allow one
to derive a stronger upper bound on $\mwimp$, despite large
uncertainties in halo WIMP velocities.

\vspace{1cm}

In this Letter, we emphasized the importance of astrophysical
uncertainties in determining the shape and overall position of the
expected detection spectrum. We did this by demonstrating how varying
the maximum of the Maxwellian velocity distribution of Galactic halo
WIMPs will considerably relax the region selected by DAMA as
corresponding to a possible annual modulation signal. (It is our
understanding that the DAMA Collaboration is currently in the process
of incorporating the effect in their annual modulation
analysis~\cite{privatecom}.) The effect will also affect WIMP
exclusion plots which are often drawn for one fixed value of $\vzero$,
and should be of interest to those involved in  searches for
halo WIMPs.

Clearly, there are several other uncertainties entering the halo WIMP
detection analysis. We reiterate that, while there is much convincing
astrophysical evidence for the existence of the Galactic halo,
parameters describing its shape have only been established through
indirect methods and their values should be considered as rather
uncertain. There still remains a considerable spread among possible
models for the Galactic halo and (halo model dependent) quoted error
bars for $\vzero$, and the local halo density should, in our opinion,
be taken with extreme caution. (In Ref.~\cite{kk98} several halo
models were compared with the standard non-rotating spherical model
and their effect on the direct rates was found to be minimal.)
Nuclear physics uncertainties in the form factors are also rather
poorly known and have been of much concern~\cite{cosmo97-spooner}.

We conclude that, due to astrophysical input uncertainties, caution
should be taken when considering exclusion regions from detection
experiments and also in analyzing their implications on underlying
WIMP models. For example, it has been claimed~\cite{bdfs98,an99} that,
in supersymmetric models with the lightest neutralino as the stable
superpartner and a WIMP candidate, it is possible to find SUSY
configurations consistent with the possible annual modulation signal
of DAMA but only for large enough $\tan\beta\gsim10$, typically rather
small relic abundance,  and for
restricted ranges of other SUSY parameters.  Including the effect of
astrophysical uncertainties  will lead to much broader ranges of
allowed parameters than previously thought~\cite{brinprep}.

\section*{Acknowledgements}
One of us (LR) is grateful to R.~Bernabei for a number of clarifying
discussions about DAMA's analysis and to T.~Sloan for helpful remarks
on the manuscript.

%

\def\NPB#1#2#3{{Nucl.\ Phys.}\/ {\bf B#1}, #3 (19#2)}
\def\PLB#1#2#3{{Phys.\ Lett.}\/ {\bf B#1}, #3 (19#2)}
\def\PRD#1#2#3{{Phys.\ Rev.}\/ {\bf D#1}, #3 (19#2)}
\def\PRL#1#2#3{{Phys.\ Rev.\ Lett.}\/ {\bf #1} #3, (19#2)}
\def\PRT#1#2#3{{Phys.\ Rep.}\/ {\bf#1}, #3 (19#2)}

\begin{thebibliography}{99}     
\singlespaced

\bibitem{jkg} G. Jungman, M. Kamionkowski and K. Griest, \PRT{267}{96}{195}.
\bibitem{annualmodulation86} A.~Drukier, \etal, Phys. Rev. {\bf D33},
3495 (1986).
\bibitem{annualmodulation88} 
K.~Freese, \etal, Phys. Rev. {\bf D37}, 3388 (1988).
\bibitem{dama98two} R. Bernabei, \etal, 
University of Rome preprint ROM2F/98/34 (27 August 1998).
\bibitem{dama98one} R. Bernabei, \etal, \PLB{424}{98}{195}.
\bibitem{dama96} R. Bernabei, \etal, \PLB{389}{96}{757}.
\bibitem{gw} M. Goodman and E. Witten, \PRD{31}{85}{3059}.
\bibitem{drees} M. Drees and M. Nojiri, \PRD{47}{93}{4226} and \PRD{48}{93}
{3483}. 
\bibitem{bb} H. Baer and M. Brhlik, \PRD{57}{98}{567}.
\bibitem{engel} J. Engel, \PLB{264}{91}{114}.
\bibitem{io} F. Iachello, L.M. Krauss and G. Maino, \PLB{254}{91}{220}.
\bibitem{na} R. Flores and J. Ellis, \NPB{400}{93}{25}.
\bibitem{quench}K. Fushimi, \etal, Phys.\ Rev.\/ {\bf C47} (1993) R425;
P.F. Smith, \etal, \PLB{379}{96}{299}; 
G.J. Davies, \etal, \PLB{322}{94}{159}. 
\bibitem{rotvel1} G.R.~Knapp, S.D.~Tremaine, and J.E.~Gunn,
Astron. J. {\bf 83} (1978) 1585.
\bibitem{rotvel2} F.J.~Kerr and D.~Lynden-Bell,
Mon. Not. R. Astr. Soc. {\bf 221} (1986) 1023.
\bibitem{rotvel3} J.A.R.~Caldwell and J.M.~Coulson, Astron. J. {\bf 93}
(1987) 1090.
\bibitem{kk98} M.~Kamionkowski and A.~Kinkhabwala, \PRD{57}{98}{3256}.
\bibitem{cosmo98-lr} L.~Roszkowski, talk at COSMO-98, Asilomar, USA,
November '98, hep-ph/9903467.
\bibitem{fornengo-idm98} N.~Fornengo, talk at IDM-98, Buxton, UK, September
'98, hep-ph/9812210.
\bibitem{privatecom} A.~Bottino and R.~Bernabei, private communication.
\bibitem{cosmo97-spooner} See, \eg, Ref.~\cite{jkg} or N.J.C.~Spooner,
Proc. Int. Workshop on Particle Physics and Cosmology (COSMO-97), World
Scientific, Ed. L.~Roszkowski.
\bibitem{bdfs98} A. Bottino, F.~Donato, N.~Fornengo, and S.~Scopel,
hep-ph/9808456, hep-ph/9808459, and hep-ph/9809239. 
\bibitem{an99} R.~Arnowitt and P.~Nath, hep-ph/9902237.
\bibitem{brinprep} M.~Brhlik and L.~Roszkowski, in preparation.
\end{thebibliography}
\end{document}